\documentclass[runningheads]{llncs}

\usepackage[T1]{fontenc}        
\usepackage{graphicx}           
\usepackage{color,soul}         
\usepackage[noadjust]{cite}
\usepackage{algorithm}          
\usepackage{algpseudocode}      

\usepackage{amsmath}            
\usepackage{amsfonts}
\usepackage{booktabs}           
\usepackage{multirow}           
\usepackage{tabularray}
\usepackage{url}

\usepackage[breaklinks]{hyperref}
\PassOptionsToPackage{unicode}{hyperref}
\PassOptionsToPackage{naturalnames}{hyperref}
\usepackage{microtype}
\usepackage[utf8]{inputenc}     
\usepackage[british]{babel}     
\usepackage{subfiles}           
\usepackage{wrapfig}            
\usepackage{xcolor}             

\usepackage{bm}                 
\usepackage{tablefootnote}      
\makeatletter
\newcommand*{\rom}[1]{\expandafter\@slowromancap\romannumeral #1@}
\makeatother
\definecolor{light-gray}{gray}{0.98}

\usepackage{caption}
\usepackage{subcaption}

\begin{document}
\title{Reconstructing Spatiotemporal Data with C-VAEs}
\author{Tiago F. R. Ribeiro\inst{1}\orcidID{0000-0003-1603-1218} 
\and 
Fernando Silva\inst{1}\orcidID{0000-0001-9335-1851} 
\and
Rog\'erio Lu\'is de C. Costa
\inst{2}\orcidID{0000-0003-2306-7585}
}
\authorrunning{T. R. Ribeiro \emph{et al}.}

\institute{CIIC, ESTG, Polytechnic of Leiria, Portugal \and CIIC, Polytechnic of Leiria, Portugal \\
\email{\{tiago.f.ribeiro, fernando.silva, rogerio.l.costa\}@ipleiria.pt}}

\maketitle              
\begin{abstract}

The continuous representation of spatiotemporal data commonly relies on using abstract data types, such as \textit{moving regions}, to represent entities whose shape and position continuously change over time. Creating this representation from discrete snapshots of real-world entities requires using interpolation methods to compute in-between data representations and estimate the position and shape of the object of interest at arbitrary temporal points. Existing region interpolation methods often fail to generate smooth and realistic representations of a region's evolution. However, recent advancements in deep learning techniques have revealed the potential of deep models trained on discrete observations to capture spatiotemporal dependencies through implicit feature learning.

In this work, we explore the capabilities of Conditional Variational Autoencoder (C-VAE) models to generate smooth and realistic representations of the spatiotemporal evolution of moving regions. We evaluate our proposed approach on a sparsely annotated dataset on the burnt area of a forest fire. We apply compression operations to sample from the dataset and use the C-VAE model and other commonly used interpolation algorithms to generate in-between region representations. To evaluate the performance of the methods, we compare their interpolation results with manually annotated data and regions generated by a U-Net model. We also assess the quality of generated data considering temporal consistency metrics.

The proposed C-VAE-based approach demonstrates competitive results in geometric similarity metrics. It also exhibits superior temporal consistency, suggesting that C-VAE models may be a viable alternative to modelling the spatiotemporal evolution of 2D moving regions.

\keywords{Region Interpolation Problem \and Moving Regions \and Conditional Variational Autoencoder \and Continuous Representation.}
\end{abstract}

\section{Introduction}


Spatiotemporal data describe phenomena, storing the time of occurrence and data on the position, shape, or dimensions of entities of interest. It is commonly used to monitor and analyze changes in land cover, optimize transportation routes, and track iceberg movement or volcanic eruptions, for example. 
Typically, such data is stored using discrete snapshots, associating a timestamp to some representation of the entity's shape and position. Nevertheless, some applications benefit from using a continuous representation of the spatiotemporal evolution of modelled entities. 

\begin{figure}[!th]
    \centering
    \includegraphics[width=1\textwidth]{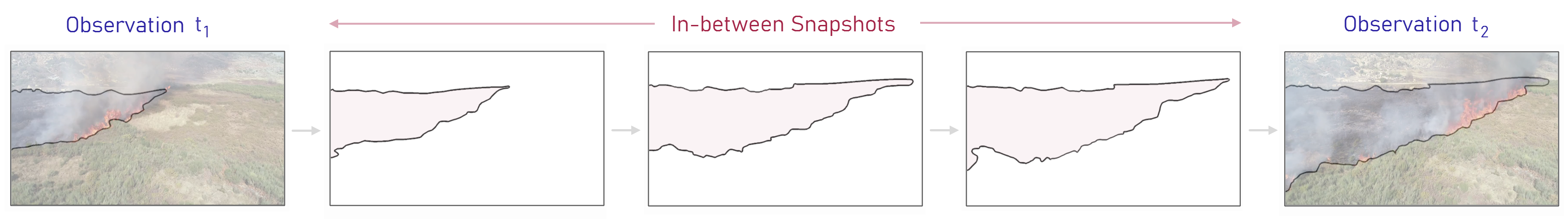}
    \caption{Continuous representation model requires a method to recreate the spatiotemporal evolution of a region, such as the progression of the burned area.}
    \vspace{-1.0em}
    \label{fig:RIP}
\end{figure}

The continuous representation frequently relies on abstract data types, such as \textit{moving regions} and \textit{moving lines}, and associates discrete representations of the modelled entities to functions that represent their evolution \cite{Tossebro_Guting_2001, rip_methods_Mckennney_2015}. It has some advantages over the discrete model, such as compression capabilities, but also has its own challenges, as creating the continuous representation of an entity requires the specification of a method to generate the entity's shape and position in-between representations. For instance, in Figure \ref{fig:RIP}, the first and last snapshots are actual images of a real-world phenomenon (forest fire burnt area), and intermediary polygons recreate the spatiotemporal evolution of the burnt area. Commonly, such a recreation employs region interpolation functions, but current methods often fail to create realistic representations of the evolution of 2D regions~\cite{Duarte_2019}. On the other hand, recent works in deep learning field have proved the capability of such models to learn from implicit features. 

In this work, we use Conditional Variational Autoencoders (C-VAEs) \cite{cVAE_2015} to create representations of the spatiotemporal evolution of 2D regions. Variational Autoencoders (VAEs) \cite{AE_Hinton} are neural networks consisting of encoders and decoders that learn the probabilistic distribution of the latent space (\emph{\emph{i.e.}} the low-dimensional representation generated by the encoder). C-VAEs extend VAEs by having conditioning information on the latent space \cite{cVAE_2015}. We assess our methods' capabilities of generating realistic representations by using them to track the dynamics of the propagation of forest fires, generating in-between observations from a discrete set of snapshots of a controlled fire captured with by stationary drone. We compare our model to two region interpolation methods from the literature (\cite{shape_based_schenk,Pyspatiotemporalgeom}), considering distinct scenarios and compression methods, and use geometric similarity metrics to compare generated polygons with ground truth data (manually annotated and automatically segmented). We also evaluate the quality of generated geometries by using temporal consistency metrics. 

Hence, the contributions of this work include (\rom{1}) employing C-VAE neural network to generate the spatiotemporal representation of real-world data; (\rom{2}) introducing a specific Temporal Consistency metric to validate the generated in-between observations; (\rom{3})  a systematic evaluation of the proposals considering distinct compression methods, and geometric similarity and temporal consistency metrics and (\rom{4}) a comparison with interpolation methods from the literature.

\section{Background and Related Work}\label{background}

Spatiotemporal data is often discretely encoded, with spatial attributes such as location and shape associated with time instants. A typical continuous representation of real-world spatiotemporal phenomena data is through \emph{moving regions}. Formally, a \emph{region} is a set of non-intersecting line segments connecting a distinct set of points and forming a closed loop that represents the external faces of a polygon. A region may contain holes, which are also delimited by line segments in a closed loop and, crucially, do not intersect the external faces~\cite{Tossebro_Guting_2001}. \emph{Moving regions} are abstract data types used to describe real-world the spatiotemporal evolution of objects of interest, \emph{i.e.}., how their shape and position change over time \cite{Forlizzi_2000}. \emph{Moving regions} are described as a series of sequentially stored regions (\emph{interval regions}) such that an interval region represents the movement of an object over a time interval between two defined instants, called \emph{slices} \cite{Tossebro_Guting_2001}.

\subsection{Moving Regions Evolution Representation}

T{\o}ssebro \emph{et al.} \cite{Tossebro_Guting_2001} proposed a framework in which \emph{moving regions} can be represented from observations stored in spatiotemporal databases. This work was, since then, expanded by different authors using the same principles \cite{rip_Heinz_Guting_2020, rip_methods_Mckennney_2015, rip_methods_Duarte_2023}. The primary objective is to produce representations that continuously maintain topological validity while ensuring consistency with the underlying spatiotemporal database systems. The \emph{Region Interpolation Problem} (RIP) is the challenge of creating a \emph{moving region} from a set of observations. Specifically, considering two observations at instants $t_1$ and $t_2$, the objective is to identify some interpolating function $f$ capable of generating a valid representation of the moving object, its position and shape at any time point between $t_1$ and $t_2$~\cite{Duarte_2019}.

Other interpolation approaches have also been suggested. For instance, if we regard a \emph{moving region} as a polyhedron, where time takes the place of a third dimension (height)~\cite{rip_Heinz_Guting_2020}, techniques used to interpolate polygons representing sections of a volumetric object, such as human organs in tomographic imaging, may be adapted to generate in-between regions. The so-called \emph{Shape-Based} interpolation is an example of one such algorithm. Contrary to the abovementioned methods, typically, this algorithm operates with raster data \cite{shape_based_schenk, shape_based_herman}. Generally, the process can be described in a sequence of steps, as follows. Let $x_1$ and $x_2$ be the 2D snapshots that contain the shape of the region at instants $t_1$ and $t_2$. For each selected snapshot, a binary image $y_k, k \in \{1,2\}$ is generated by segmenting the region of interest. Next, a grey-level distance map $z_k, k \in \{1,2\}$ is generated for each binary image $y_k$ by mapping the Euclidean distance to the boundary of the region. The distance values inside the region are set to positive and the distance values outside the region are set to negative. The maps $z_k$ are then reconstructed by using linear interpolation at the pixel level. The shape of the region at a given arbitrary point in time $t_i, t_1 < t_i < t_2 $ is found by identifying the zero-crossings of the interpolated distance maps. Finally, these contours generate the region of interest in $t_i$.

An alternative strategy to interpolating regions involves applying algorithms that can learn the phenomena representations. Deep learning models have shown promising results in various image interpolation applications. For instance, in~\cite{AE_interpol}, Oring \emph{et al.} explore interpolation techniques for raster images depicting polyhedra at different angles and other geometric representations with deformable objects, by \emph{smoothly} interpolating the latent space of Autoencoder models. Similarly, Cristovao \emph{et al.} and Mi \emph{et al.} propose equivalent methods for interpolating raster images with three-dimensional objects at various angles, as well as snapshots representing moving objects, using the latent space interpolation of various Generative Latent Variable models~\cite{in_between_imgs_VAE,latent_space_interpol_LU}. 
In this work, we apply C-VAEs to generate in-between representations of the evolution of 2D regions.

\subsection{Conditional Variational Autoencoders}

An \textbf{Autoencoder} (AE)~\cite{AE_Hinton} is a neural network that takes a high-dimensional input $\mathbf{x}\in\mathbb{R}^D$, such as an image, and maps it to a compact, low-dimensional representation $\mathbf{z}\in\mathbb{R}^d$, referred to as the \emph{latent space} $z$, which is typically a vector. This compressed representation is then used by a decoder to reconstruct the original input. The AE architecture can be decomposed into three components: the encoder $f_{\phi}(\cdot)$, which maps the input to the latent space, a decoder $g_{\theta}(\cdot)$, which maps the latent representation back to input space, and a bottleneck $\mathbf{z}$ that stores the compressed codes. The encoder and decoder are often implemented as neural networks with learnable parameters $\phi$ and $\theta$, respectively.

The \textbf{Variational Autoencoder} (VAE)~\cite{VAE_Kingma2013, VAE_Rezende2014}, an AE development, consist of an encoder $q_{\phi}(z|x)$ and a decoder $p_{\theta}(x|z)$. Unlike regular AEs, VAEs learn a probabilistic distribution of the latent space instead of a deterministic mapping. VAEs are trained to minimize the evidence lower bound (ELBO) on $\log p(x)$, where $p(x)$ is the data generating distribution. The ELBO can be expressed as: 
\begin{equation}
\mathcal{L}(\theta, \phi, x) = \mathbb{E}{q{\phi}(z|x)}[\log p_{\theta}(x|z)] - D_{KL}(q_{\phi}(z|x) || p(z))
\end{equation}

Here, $p(z)$ is a chosen prior distribution, such as a multivariate Gaussian distribution with mean zero and identity covariance matrix. The encoder predicts the mean $\mu_{\phi}(x)$ and standard deviation $\sigma_{\phi}(x)$ for a given input $x$, and a latent sample $\hat{z}$ is drawn from $q_{\phi}(z|x)$ using the reparameterization trick: $\hat{z} = \mu_{\phi}(x) + \sigma_{\phi}(x) * \epsilon$, where $\epsilon \sim \mathcal{N}(0,I)$. By choosing a multivariate Gaussian prior, the KL divergence term can be calculated analytically.
The first term in the ELBO equation is typically approximated by calculating the reconstruction error between many samples of $x$ and their corresponding reconstructions $\hat{x} = D_{\theta}(E_{\phi}(x))$. New samples, not present in the training data, can be synthesized by first drawing latent samples from the prior, $z \sim p(z)$, and then drawing data samples from $p_{\theta}(x|z)$, which is equivalent to passing the latent samples through the decoder, $D_{\theta}(z)$. 
The VAEs architecture enables better interpolation than traditional AEs because they learn a continuous latent space that can be easily sampled to generate new data, leading to smooth transitions in the generated outputs~\cite{Berthelot_2018_VAE_interpol}.

\textbf{Conditional Variational Autoencoders} (C-VAEs)~\cite{cVAE_2015} extend VAEs to learn a conditional distribution $p_{\theta}(x|y)$ where $y$ represents some conditioning information, such as class labels. C-VAEs consist of an encoder $q_{\phi}(z|x,y)$ and decoder $p_{\theta}(x|z,y)$, both of which take in the conditioning information $y$. C-VAEs are also trained to minimize the ELBO on $\log p_{\theta}(x|y)$.

The ELBO for C-VAEs is similar to that of VAEs, but conditioned on $y$:
\begin{equation}
\mathcal{L}(\theta, \phi, x, y) = \mathbb{E}{q{\phi}(z|x, y)}[\log p_{\theta}(x|z, y)] - D_{KL}(q_{\phi}(z|x, y) || p(z|y))
\end{equation}
where $p(z)$ is a chosen prior distribution. The encoder predicts the mean $\mu_{\phi}(x,y)$ and standard deviation $\sigma_{\phi}(x,y)$ for a given input $(x,y)$, and a latent sample $\hat{z}$ is drawn from $q_{\phi}(z|x,y)$ as follows: $\epsilon \sim \mathcal{N}(0,I)$ then $z=\mu_{\phi}(x,y)+\sigma_{\phi}(x,y)*\epsilon$. The first term in the loss function is typically approximated by calculating the reconstruction error, such as Mean Squared Error or Binary Cross-Entropy losses, between many samples of $x$ and $\hat{x}=D_{\theta}(E_{\phi}(x,y))$.

\section{Spatiotemporal Data Compression and Reconstruction}\label{methodoly}

By capturing the evolution of real-world objects through images or videos, for example, we are generating a sample of the space-time evolution of an entity. This discrete sample leads to a compressed representation, but also to data loss. For the reconstruction of such data, a method capable of representing the evolution of the object of interest is required~\cite{Mckenney_mov_regions}. Several factors can impact the quality of the reconstruction, including the used compression technique~\cite{costa2020sampling}. 

\subsection{C-VAE Based Interpolation}\label{cvae_interpol_section}

In addition to generating new samples conditioned on $y$, C-VAEs have the ability to perform conditional image editing in the latent space. Given two different conditioning inputs, $x_1$ and $x_2$, one can interpolate in the latent space between the corresponding latent codes $z_1$ and $z_2$ to generate novel images that smoothly blend the two conditioning factors. This capability makes C-VAEs suitable to interpolate different discrete or continuous codified representations. 

Our approach consists in training a C-VAE model on the set of samples to be interpolated, conditioned by the timestamp of each sample. For applications that operate with discrete variables, \emph{i.e.} limited number of classes, it is common to encode the classes with one-hot encoding and then concatenate that vector to the input as well as the latent space.

\begin{figure}[!th]
    \centering
    \includegraphics[width=1\textwidth]{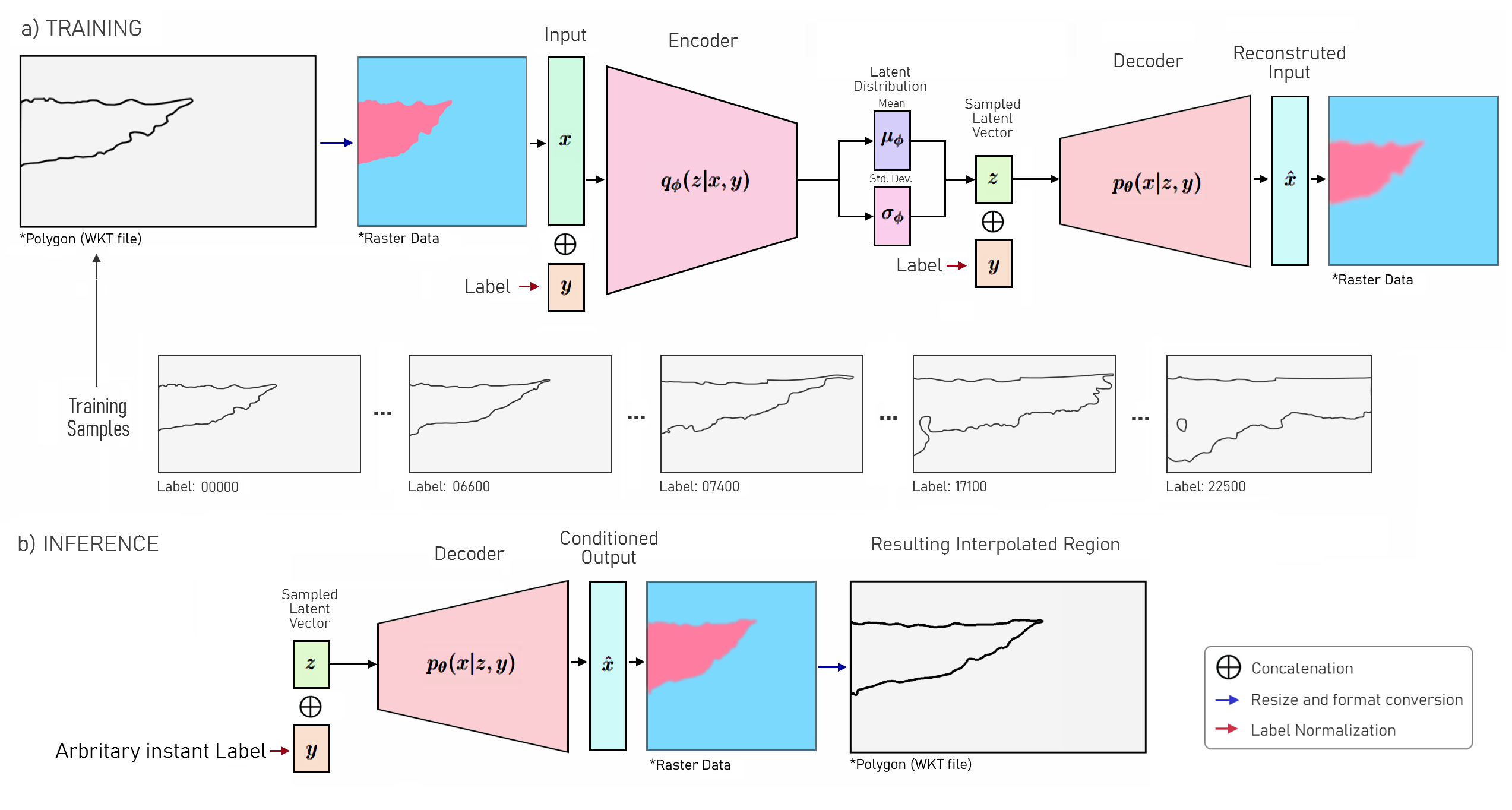}
    \caption{\textbf{Employed C-VAE Architecture.} a) each region stored in WKT format is converted to raster image to be processed by the model b) a new image is generated conditioned by a label and converted to WKT format.}
    \vspace{-1.5em}
    \label{fig:cvae_train}
\end{figure}

Since we are dealing with continuous phenomena, and the conditioning variable representing the interpolating instant (\texttt{label}) is continuous, we chose to simply use the frame number of the original video, normalize it to a range between 0 and 1, and then concatenate the resulting value $y$ to the latent space $z$ and the input $x$, as shown in Figure \ref{fig:cvae_train}. Other ways of encoding the timestamp could be considered. Still, for this specific application, our approach proved to be sufficient, even considering the introduction of the quantization error related to the finite resolution of the floating-point numbers used to store the latent space values.

During the inference stage, we sample both the latent space variable $z$ and the conditioning variable $y$ to generate new samples from the learned conditional distribution $p_{\theta}(x|y)$. More specifically, we first sample a random vector $\epsilon$ from a standard normal distribution, then use it to compute a sample $z$ from the learned approximate posterior distribution $q_{\phi}(z|x,y)$ using the reparameterization trick. Next, we define a specific conditioning variable $y_i$, representing an arbitrary instant within the length of the video, and concatenate it with the sampled $z$ to the decoder network $p_{\theta}(x|z,y)$ to generate a new sample $\hat{x}_i$, an estimated raster image for the instant $i$.



\subsection{Compression Methods}\label{comp_methods}


The sampling strategy reduces the amount of spatiotemporal data stored but also determines the representation used for interpolation, thus impacting the performance of reconstruction methods \cite{costa2020sampling}. 

\medskip

\noindent \textbf{Periodic Sampling.}\label{period_sampl} As a first compression approach, we consider the regions corresponding to the burned area as a sequence of observations $x_t = \{x_1, x_2, ..., x_n\}$, where $n$ is the length of the sequence, ordered in time with a given sampling frequency $f_s$, corresponding to the video frame rate. Then we sample the sequence periodically using some downsampling factor $d \in \mathbb{N}$. This results in a new sequence of observations $w_t = \{w_1, w_2, ..., w_m\}$, where $m = \lfloor n/d \rfloor$ is the length of the downsampled sequence. Each observation $y_i$ corresponds to the original observation $x_{id}$, where $i$ is the index of the downsampled sequence and $d$ is the downsampling factor. This approach reduces the size of the sequence by a factor of $d$, however, this method may discard relevant samples.

\medskip

\noindent \textbf{Distance-Based Sampling.}\label{dist_sampl} As a second strategy, we follow the method suggested in \cite{costa2020sampling}. This method downsamples a sequence of observations by selecting representative points that are dissimilar from each other. It takes a set of observations and a distance function that calculates the dissimilarity between two observations, along with a threshold value $\alpha$. The algorithm initializes the downsampled sequence with the first observation, and iteratively adds subsequent observations to the sequence only if they are more dissimilar than the threshold value from the last selected observation. This process continues until all observations have been considered, or the sequence length reaches the desired length. The distance function can be any metric that calculates dissimilarity, such as Jaccard distance, Hausdorff distance, or a combination of several metrics.

\subsection{Quality Evaluation Metrics}\label{metrics}

To evaluate the performance of the spatiotemporal representation, we use two similarity metrics between generated images and ground truth data. We also propose a temporal consistency metric that measures the consistency between generated representations considering phenomena-specific features.

\medskip

\noindent \textbf{Jaccard Index}. The Jaccard Index (JI) measures the overlap or similarity between two shapes or polygons ($A$ and $B$). It returns a value between 0 and 1, where 1 means the shapes are identical and 0 means they have no overlap:


\begin{equation}\label{JI}
    JI(A,B) = \frac{A\cap B}{A\cup B}
\end{equation}


\smallskip

\noindent \textbf{Hausdorff Distance}. The Hausdorff distance measures the degree of mismatch between two non-empty sets of points $A = \{a_1,...,a_p\}$ and $B = \{b_1,...,b_p\}$ by measuring the distance of the point $A$ that is farthest from any point of $B$ and vice versa~\cite{hausdorff_distance}. In simpler terms, it measures how far apart two sets are from each other by finding the maximum distance between a point in one set and its closest point in the other set. 
Formally, it can be denoted as follows:
\begin{equation}\label{HD}
    HD(A,B) = max(h(A,B), h(B,A))
\end{equation}
where $h(A,B)$ is the directed Hausdorff distance from shape $A$ to shape $B$, and $h(B,A)$ is the directed Hausdorff distance from shape $B$ to shape $A$. The directed Hausdorff distance is defined as: 

\begin{equation}\label{h_ab}
    h(A,B) = \underset{a \in A}{max}\:\underset{b \in B}{min}|| a - b ||
\end{equation}
where $a$ is a point in shape $A$, $b$ is its closest point in shape $B$, and $||a - b||$ denotes the Euclidean distance between two points.


\medskip

\noindent \textbf{Temporal Consistency}. We know that for the same fire outbreak, an area that is established as burned cannot cease to be so at a later stage. Likewise, we know that the burned area never decreases. With these considerations in mind, we define the temporal consistency $TC$ as a complement of a geometric difference:
\begin{equation}\label{TI}
    TC_{stride} = 1 -  \frac{A_{t} - A_{t+stride}}{A_{t+stride}},  \forall t \in \{1, 2, ... ,T-stride\}
\end{equation}
where $A_{t}$ and $A_{t+stride}$ represent the burned area region in separated by $stride$ samples. To assess different time scales, we consider various values of stride from a geometric progression $stride_n = ar^{n-1}$, $\forall n \in \{1, 2, ... ,N\}$, with $N$ smaller than the total number of polygons in the sequence, $a$ the coefficient of each term and $r$ is the common ratio between adjacent terms. 
To assess the performance for each stride, we calculate the average of $TC_{stride}$. Finally, we can also estimate the overall Temporal Consistency by computing the mean of all $TC_{stride}$ averages.

\section{Performance Evaluation}\label{experiments}



Our study uses data from a video captured by a DJI Phantom 4 PRO UAV equipped with an RGB camera during a prescribed fire at Torre do Pinh\~ao, in northern Portugal (41° 23’ 37.56", -7° 37’ 0.32"). The UAV remained in a nearly stationary stance during the data collection. The footage is approximately 15 minutes long, with a frame rate of 25 fps and a resolution of 720$\times$1280, amounting to 22500 images. Prior works have employed this video, as referenced in~\cite{rog_costa_burned_3,rog_costa_burned_4}.

\begin{figure}[ht!]
    \centering
    \includegraphics[width=1\textwidth]{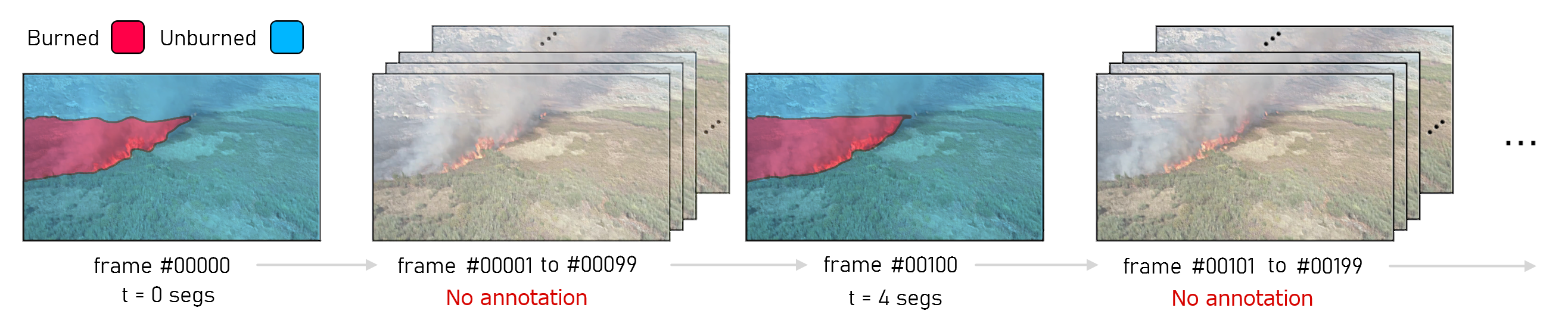}
    \caption{\textbf{\emph{BurnedAreaUAV} dataset.} The annotation was done for the entire length of the video for \texttt{burned\_area} and \texttt{unburned\_area} classes.}
    \vspace{-1.5em}
    \label{fig:annotations}
\end{figure}


In~\cite{artigo_seg_burned}, we manually annotated the burned area, as outlined in Figure \ref{fig:annotations}. The process yielded 249 annotated frames, of which 226 were used for training and 23 regions for testing. The training and testing sets were created with a periodicity of 100 frames, but having an offset of 50 frames between them. This training subset is considered a reliable representation of the polygon representing the burnt area and serves as the basis for interpolation with \emph{Periodic Sampling}. We further subsample the resulting 226 frames set by applying Distance-Based Sampling, using Jaccard's distance as a dissimilarity measure and a tolerance threshold of $\alpha = 0,15$, compressing the number of samples to 13.








Additionally, we trained a segmentation model based on the U-Net architecture~\cite{Ronneberger2015} from scratch on the \emph{BurnedAreaUAV} training set, producing segmentation masks for all video frames, which were then converted to Well-known Text (WKT) compatible polygons. The polygons produced by the U-Net model obtained an overall Jaccard index value superior to 0.95 on the test set of the \emph{BurnedAreaUAV} dataset, and are therefore considered good approximations of the actual progression of the burnt area. We designated this set of 22,500 polygons as \emph{U-Net Samples}.


\subsection{Evaluation Scenarios}

\medskip

\noindent \textbf{Experiment Description.} In this experiment we evaluated three different algorithms: (\textbf{\rom{1}}) the Mckenney interpolation method~\cite{Pyspatiotemporalgeom}, (\textbf{\rom{2}}) the Shape-Based interpolation~\cite{shape_based_schenk, shape_based_Bouazizi_2021} and (\textbf{\rom{3}}) the C-VAE-based method described in Section~\ref{cvae_interpol_section}.

For each algorithm, we generate in-between samples corresponding to the frame timestamps of the original video by employing the 226 samples resulting from Periodic Sampling as well as the subset of 13 Distance-Based samples. We compare the polygons generated by the algorithms to the ones generated by the automatic segmentation (\emph{U-Net Samples}) and validated using the test subset of the \emph{BurnedAreaUAV} dataset and calculate the Jaccard and Hausdorff similarity metrics. We also evaluate the quality of the generated polygons in terms of the Temporal Consistency indicator formulated in Section~\ref{metrics}. To calculate the JI and HD, we discarded the samples that supported the calculation of the intermediate regions, both for Periodic Sampling and for Distance-Based sampling. That is, out of a universe of 22,500 observations corresponding to the video frames, we considered 22,274 intermediate regions for the Periodic Sampling and 22,487 for the Distance-Based sampling. All the metrics were calculated considering the resolution of the original footage ($1280\times720$).

\medskip

\noindent \textbf{Experimental Setup.} The experiments were conducted on a computer running Windows 10 and equipped with an Intel i7-10700K processor, an Nvidia GeForce RTX 3090 GPU, and 32 GB of RAM. The code was developed in Python, almost entirely in Jupyter Notebooks. Our C-VAE model is built upon a typical convolutional-based Neural Network implementation for the encoder and decoder, with minor hyperparameter tuning. The code is available in  \url{https://github.com/CIIC-C-T-Polytechnic-of-Leiria/Reconstr_CVAE_paper}.

\subsection{Experimental Results} 


Table~\ref{tab:similarity} presents the achieved values for the similarity metrics and Table~\ref{tab:tc} summarizes the results in terms of the temporal consistency evaluation. 

\begin{figure}[ht!]
    \centering
    \begin{tabular}{cccc}
    \subfloat[]{\includegraphics[width=0.437\textwidth, height=0.3\textwidth]{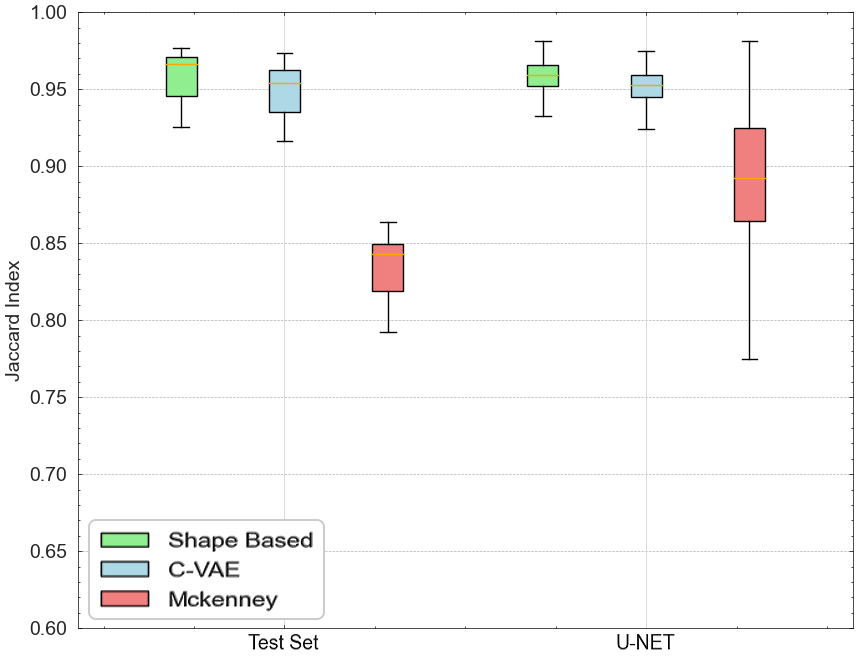}} &
    \subfloat[]{\includegraphics[width=0.437\textwidth, height=0.30\textwidth]{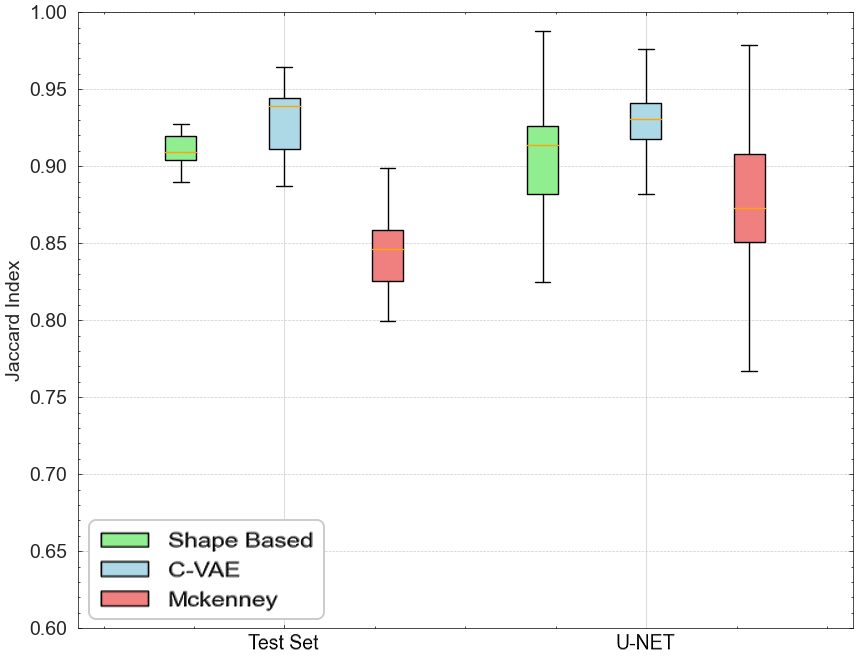}} &\\
    \subfloat[]{\includegraphics[width=0.437\textwidth, height=0.30\textwidth]{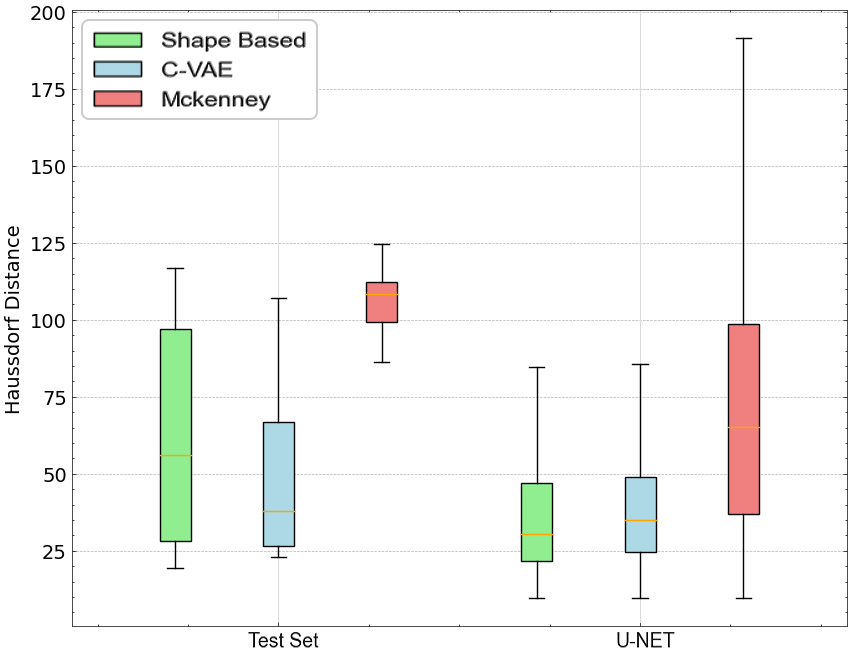}} &
    \subfloat[]{\includegraphics[width=0.437\textwidth, height=0.30\textwidth]{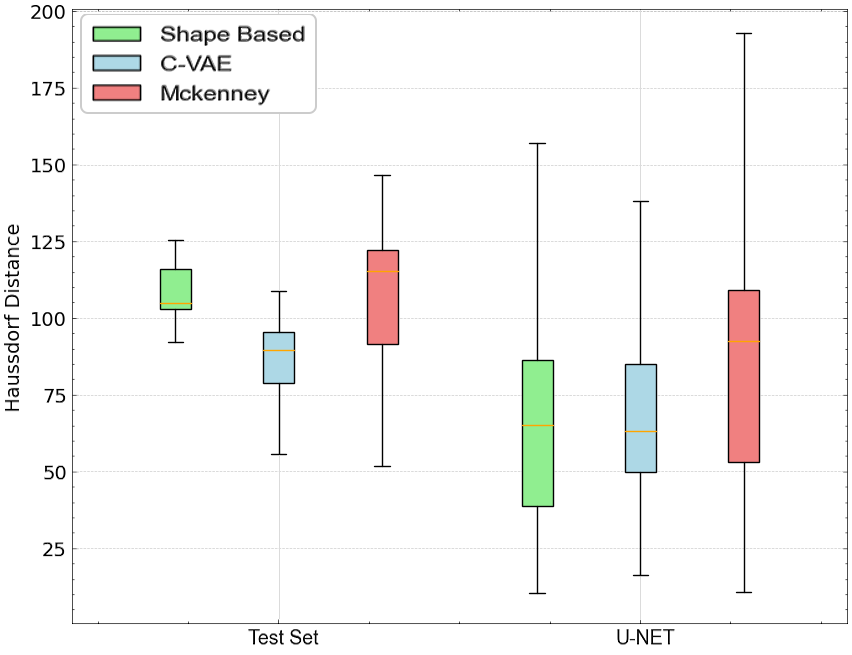}} &\\
    \end{tabular}
    \caption{Boxplots comparing the performance in terms of JI and HD metrics. (a) JI for periodic sampling, (b) JI for distance-based sampling, (c) HD for periodic sampling,  (d) HD for distance-based sampling.}
    \vspace{-1.5em}
    \label{iou_hd}
\end{figure}

The Shape-Based and C-VAE interpolations outperformed the Mckenney interpolation method on both the periodic and distance-based sampling (as shown in Table~\ref{tab:similarity} and in Figure~\ref{iou_hd} on the top). That is particularly notable on the \textit{BurnedAreaUAV} test set, on which the Shape-Based algorithm and the C-VAE algorithm show relatively close performance, having the former a small advantage.
The Shape-Based algorithm on both datasets achieved the best performance in terms of the Hausdorff Distance metric. 
Comparatively low results on the U-Net dataset 
are due to the error inherent to the auto-generated segmentation. We can also observe that reducing the number of support samples for interpolation did not have a very pronounced impact on the Jaccard Index or Hausdorff distance values, which supports the validity of distance based compressing algorithm (from 226 to 13 samples) for these particular datasets

\begin{table}[th!]
    \fontsize{8pt}{8pt}\selectfont
    \centering
    \caption{\textbf{Similarity Evaluation.} Comparison of JI and HD for U-Net Samples and \textit{BurnedAreaUAV} test set using periodic and distance-based sampling.}
    \begin{tabular}{@{}llccccccc@{}}
    \toprule
    \multicolumn{9}{c}{\textbf{\textsc{Periodic Sampling}}}\\
    \midrule
    \multirow{2}{*}{\textsc{Dataset}} & \multirow{2}{*}{\textsc{Algorithm}} & 
    \multicolumn{3}{c}{\textsc{Jaccard Distance}} & \multicolumn{1}{c}{\textsc{ }}& \multicolumn{3}{c}{\textsc{Hausdorff Distance}} \\ 
    \cmidrule(l){3-5}
    \cmidrule(l){6-9}
     & &  Mean & SD & min-max & & Mean & SD & min-max \\ 
    \cmidrule(l){0-5}
    \cmidrule(l){6-9}
    \multirow{3}{*}{\begin{tabular}[c]{@{}l@{}}U-Net Samples\end{tabular}}
     & Shape-Based & \textbf{0.958} & \textbf{0.011} & \textbf{0.870-0.982} && 42.460 & 37.503 & 9.849-243.994 \\
     & C-VAE & 0.951 & 0.011 & 0.852-0.975 && \textbf{41.866} & \textbf{26.045} & \textbf{9.849-242.745} \\
     & Mckenney & 0.892 & 0.048 & 0.519-0.982 && 72.195 & 44.284 & 9.659-364.660 \\ 
    \cmidrule(l){0-5}
    \cmidrule(l){6-9}
    \multirow{3}{*}{\begin{tabular}[c]{@{}l@{}}\textit{BurnedAreaUAV}\\ Test Set\end{tabular}} 
     & Shape-Based &\textbf{0.959} & \textbf{0.016} & \textbf{0.925-0.977} && \textbf{48.382} &  \textbf{33.312} & \textbf{19.444-117.000}  \\
     & C-VAE & 0.949 & 0.017 & 0.916-0.974 & &60.815 & 24.926 & 23.000-107.201 \\
     & Mckenney & 0.822 & 0.073 & 0.493-0.864 && 113.161 & 33.832 & 86.279-266.303  \\ 
     \end{tabular}
    \begin{tabular}{@{}llccccccc@{}}
    \toprule
    \multicolumn{9}{c}{\textbf{\textsc{Distance Based Sampling}}}\\
    \midrule
    \multirow{3}{*}{\begin{tabular}[c]{@{}l@{}}U-Net Samples\end{tabular}}
     & Shape-Based & \textbf{0.928} &  \textbf{0.020} & \textbf{0.845-0.982}  & &\textbf{68.315} & \textbf{38.443} & \textbf{10.296-306.026} \\
     & C-VAE & 0.905 & 0.026 & 0.825-0.987 & & 76.464 & 52.362 & 16.000-338.095 \\
     & Mckenney & 0.876 & 0.040 & 0.763-0.978 & & 85.426 & 38.922 & 10.471-269.194  \\  
    \cmidrule(l){0-5}
    \cmidrule(l){6-9}
    \multirow{3}{*}{\begin{tabular}[c]{@{}l@{}}\textit{BurnedAreaUAV}\\ Test Set\end{tabular}} 
     & Shape-Based  & 0.910 & 0.011& 0.889-0.928  & & \textbf{60.815} &  \textbf{33.312} & \textbf{19.444-117.000}\\
     & C-VAE & \textbf{0.930} & \textbf{0.021} & \textbf{0.887-0.964} &  &85.220 & 14.827 & 52.773-108.853  \\
     & Mckenney & 0.850 & 0.038 & 0.799-0.960 & & 103.068  & 30.744 & 23.014-146.521  \\ 
     \bottomrule
    \end{tabular}
    \vspace{-1.0em}
    \label{tab:similarity}
\end{table}

\begin{table}[th!]
    \fontsize{8pt}{8pt}\selectfont
    \centering
    \vspace{-1.5em}
    \caption{\textbf{Temporal Consistency Comparison.} Average temporal consistency across different algorithms for periodic and distance-based sampling.}
    \begin{tabular}{@{}llll|llll@{}}
    \toprule
    \multicolumn{4}{c|}{\textbf{\textsc{Periodic Sampling}}} & \multicolumn{4}{c}{\textbf{\textsc{Distance Based Sampling}}} \\
    \midrule
    \multirow{2}{*}{\textsc{Algorithm}} & \multicolumn{3}{c|}{\textsc{Avg. Temp. Consistency}} & \multirow{2}{*}{\textsc{Algorithm}} & \multicolumn{3}{c}{\textsc{Avg. Temp. Consistency}}\\
    \cmidrule(l){2-4}
    \cmidrule(l){6-8}
     &  Mean & SD & min-max &&Mean & SD & min-max \\
    \midrule
    Shape-Based & 0.986 & 0.011 & 0.971-1.000 & Shape-Based & 0.994 & 0.006 & 0.985-1.000 \\
    C-VAE & \textbf{0.993} & \textbf{0.007} & \textbf{0.982-0.998} & C-VAE & \textbf{0.999} & \textbf{0.001} & \textbf{0.997-1.000} \\
    Mckenney & 0.970 & 0.019 & 0.951-0.995 &Mckenney & 0.983 & 0.018 & 0.948-0.998 \\
    \bottomrule
    \end{tabular}
    \vspace{-1.5em}
    \label{tab:tc}
\end{table}


\begin{figure}[th!]
    \centering
    \begin{tabular}{cc}
    \subfloat[Periodic Sampling]{\includegraphics[width=0.95\textwidth, height =0.3\textwidth]{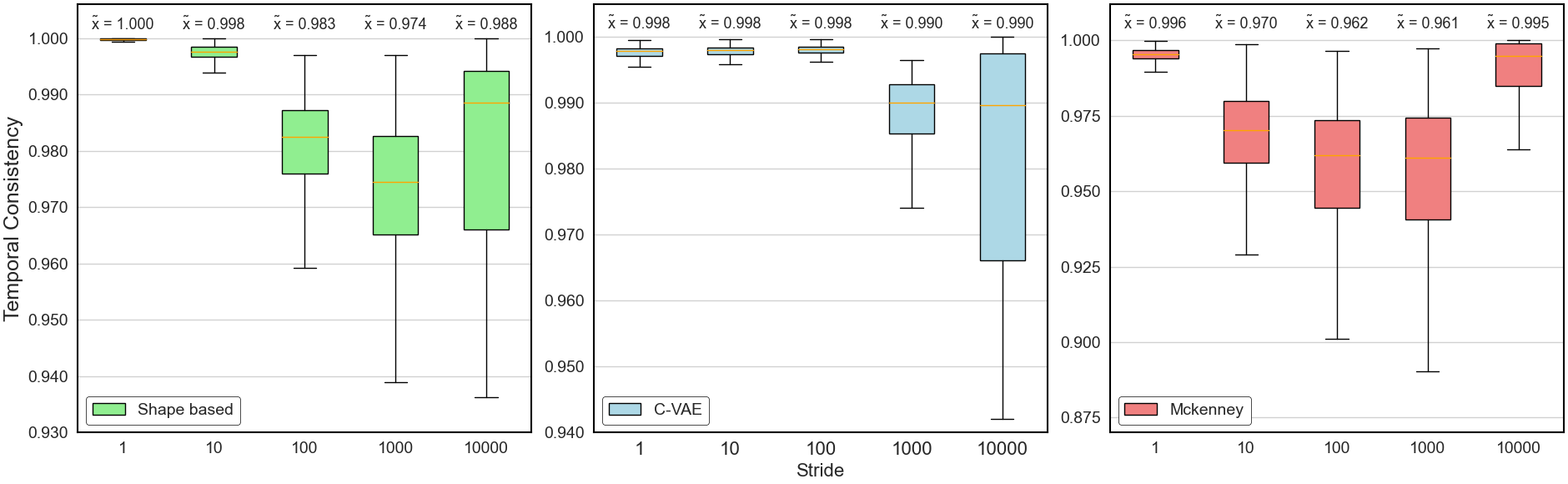}}\\
    \subfloat[Distance Based Sampling]{\includegraphics[width=0.95\textwidth, height =0.3\textwidth]{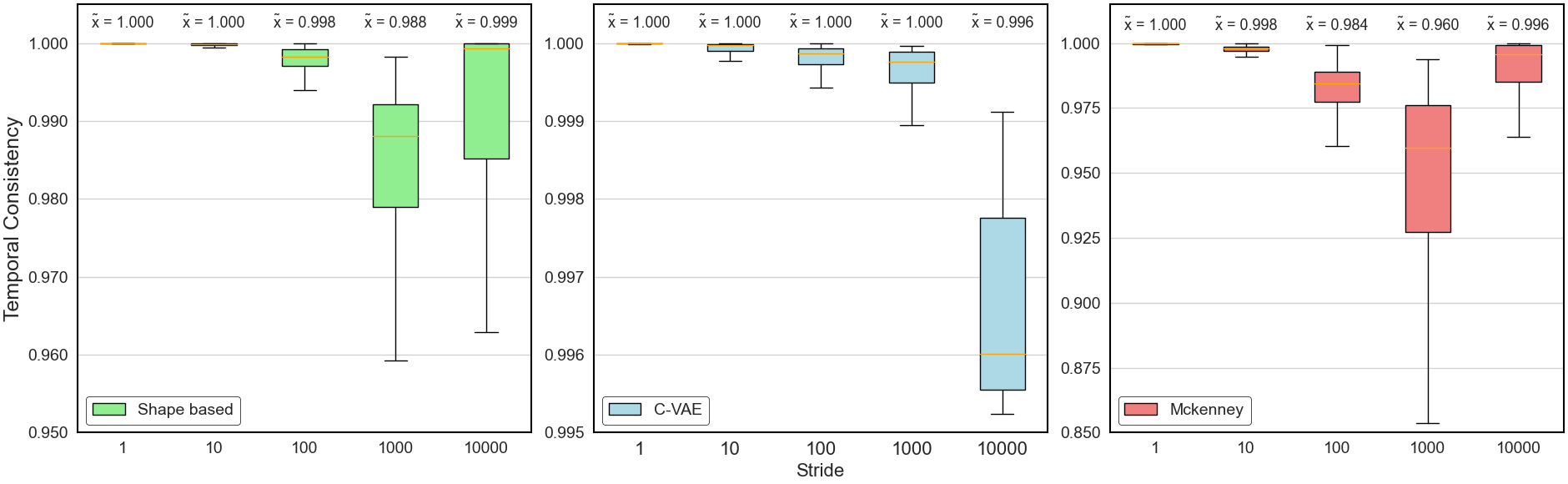}}\\
    \end{tabular}
    \caption{Results for Periodic Sampling and Distance Based Sampling, and different temporal strides. Different y-axis scales are used for better visibility.}
    \vspace{-0.5em}
    \label{fig:TC}
\end{figure}


\begin{figure}[th!]
    \centering
    \includegraphics[width=1\textwidth, height =0.28\textwidth]{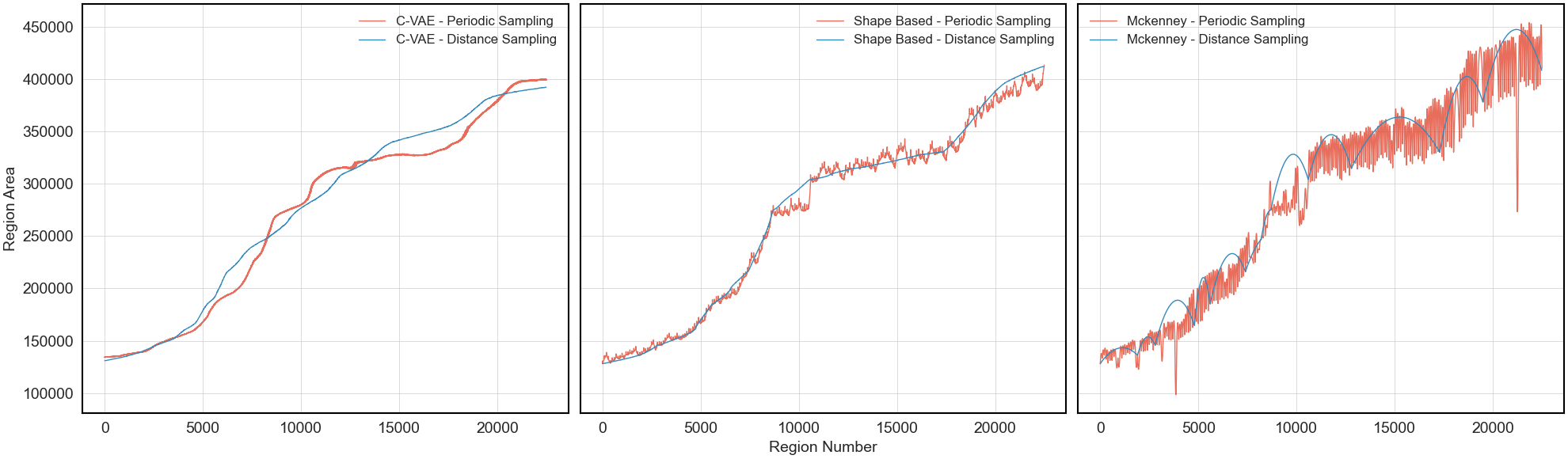}
    \caption{Representation of the evolution burned area.}
    \vspace{-1.5em}
    \label{fig:area}
\end{figure}

The results in Table~\ref{tab:tc} indicate the average Temporal Consistency for all considered stride values (1, 10, 100, 1,000 and 10,000) and show that the C-VAE model can produce polygons with higher consistency for the burnt area evolution phenomenon in both datasets. Figure~\ref{fig:area} corroborates that by showing the superior monotonicity and \emph{smoother} evolution of the burned area representations generated by the C-VAE model. Analysis of Figure~\ref{fig:TC} indicates that the C-VAE algorithm is superior for strides up to 1,000 and performs less well for strides of 10,000, which indicates less ability to retain consistency over a longer time window.

\subsection{Discussion}

A high-quality interpolation should exhibit two main features: firstly, the points lying between the interpolated values should be virtually identical to the ground truth data; secondly, the intermediate points should enable a smooth, semantically coherent transition between the endpoints.
As McKenney's algorithm focuses on creating interpolations with valid topology (i.e. without self-intercept segments), the sequences of generated regions in a real-world, noisy dataset like the \emph{BurnedAreaUAV}, revealed deformations and incoherences. 
Other authors (e.g., \cite{moreira_2016, Duarte_2018}) also referred to these issues, reflected in our results. 

Although the Shape-Based algorithm achieved a good similarity score, it tends to produce artefacts and struggles to interpolate polygons with significantly different topologies~\cite{Lee_2002}. The lower performance observed for interpolations using distance-based sampling in the \emph{BurnedAreaUAV} test subset may reflect this issue, but additional experimentation is needed to confirm this hypothesis.

C-VAEs can learn continuous and smooth representations of complex high-dimensional data by simultaneously optimizing two loss terms: the reconstruction and the KL divergence losses. It is encouraged to describe the latent state for an observation with distributions close to the prior, but deviate when necessary to describe the input's salient features. Minimizing the KL divergence forces the encodings to be close to each other while still remaining distinct, allowing for smooth interpolation and the construction of new samples. In other words, the equilibrium reached by the cluster-forming nature of the reconstruction loss and the dense packing nature of the KL loss causes the formation of distinct clusters, enabling the decoder to interpolate smoothly and avoiding sudden gaps between clusters. This mechanism explains the superior Temporal Consistency achieved by the proposed C-VAE-based solution. 
However, a C-VAE also has its drawbacks. First, unlike the classical models, it has to be trained before the interpolation, which may be time-consuming. Secondly, standard VAEs tend to generate blurry outputs, translating into a poor definition of the region boundaries.
 
\section{Conclusion}\label{conclusion}



The continuous representation of spatiotemporal data requires methods for generating an entity representation for in-between observations. To implement the \textit{moving regions} abstraction, these methods are usually region interpolation algorithms. But the recent advances in deep models show they may be a possible alternative to this problem.

In this work, we compare the performance of a C-VAE deep model with classical algorithms for the task of region interpolation. We use two datasets obtained from an aerial video of a prescribed fire and assess the performance of the solutions using geometric similarity and temporal consistency metrics.

The C-VAE algorithm performed competitively against the best-performing algorithm (Shape-Based) in terms of similarity metrics and also achieved superior temporal consistency. The results suggest that VAE-based models are viable options for spatiotemporal interpolation and motivate us to explore different AE variants to address the identified limitations. The C-VAE model generated a relatively realistic and smooth representation of the phenomenon evolution, a challenge faced by region interpolation methods.

In the future, we plan to test AE-based models' capabilities to generate the spatiotemporal evolution of a wider range of real-world phenomena. 

\section{Data availability}

The source code is available at the following \url{https://github.com/CIIC-C-T-Polytechnic-of-Leiria/Reconstr\_CVAE\_paper}. Additionally, the dataset can be accessed directly from this link \url{https://zenodo.org/record/7944963/#.ZGYP6nbMIQ8}.

\section{Declaration of Competing Interest}

The authors declare that they have no known competing financial interests or personal relationships that could influence the work reported in this paper.

\section{Acknowledgements}

This work is partially funded by FCT - Funda\c{c}\~ao para a Ci\^encia e a Tecnologia, I.P., through projects MIT-EXPL/ACC/0057/2021 and UIDB/04524/2020, and under the Scientific Employment Stimulus - Institutional Call - CEECINST/00051/2018.

This preprint has not undergone peer review or any post-submission improvements or corrections. The Version of Record of this contribution is published in Advances in Databases and Information Systems. ADBIS 2023. Lecture Notes in Computer Science, vol 13985. Springer, and is available online at \url{http://dx.doi.org/10.1007/978-3-031-42914-9_5}.

\bibliographystyle{splncs04}
\bibliography{refs.bib}       

\begin{thebibliography}{10}
\providecommand{\url}[1]{\texttt{#1}}
\providecommand{\urlprefix}{URL }
\providecommand{\doi}[1]{https://doi.org/#1}

\bibitem{Berthelot_2018_VAE_interpol}
Berthelot, D., Raffel, C., Roy, A., Goodfellow, I.J.: Understanding and
  improving interpolation in autoencoders via an adversarial regularizer. CoRR
  \textbf{abs/1807.07543} (2018)

\bibitem{shape_based_Bouazizi_2021}
Bouazizi, K., Zarai, M., Dietenbeck, T., Aron-Wisnewsky, J., Clément, K.,
  Redheuil, A., Kachenoura, N.: Abdominal adipose tissue components
  quantification in mri as a relevant biomarker of metabolic profile. Magnetic
  Resonance Imaging  \textbf{80},  14--20 (2021)

\bibitem{costa2020sampling}
Costa, R.L.C., Miranda, E., Dias, P., Moreira, J.: Sampling strategies to
  create moving regions from real world observations. In: Proceedings of the
  35th Annual ACM Symposium on Applied Computing (ACM SAC). pp. 609--616 (2020)

\bibitem{rog_costa_burned_3}
Costa, R.L.C., Miranda, E., Dias, P., Moreira, J.: Experience: Quality
  assessment and improvement on a forest fire dataset. J. Data and Information
  Quality  \textbf{13}(1) (2021)

\bibitem{rog_costa_burned_4}
Costa, R.L.C., Moreira, J.: Automatic quality improvement of data on the
  evolution of 2d regions. In: Advanced Data Mining and Applications. pp.
  288--300 (2022)

\bibitem{in_between_imgs_VAE}
Cristovao, P., Nakada, H., Tanimura, Y., Asoh, H.: Generating in-between images
  through learned latent space representation using variational autoencoders.
  IEEE Access  \textbf{8},  149456--149467 (2020)

\bibitem{Duarte_2019}
Duarte, J., Silva, B., Moreira, J., Dias, P., Miranda, E., Costa, R.L.C.:
  Towards a qualitative analysis of interpolation methods for deformable moving
  regions. p. 592–595. SIGSPATIAL '19 (2019)

\bibitem{Duarte_2018}
Duarte, J., Dias, P., Moreira, J.: An Evaluation of Smoothing and Remeshing
  Techniques to Represent the Evolution of Real-World Phenomena: 13th
  International Symposium, ISVC 2018, Las Vegas, NV, USA, November 19 – 21,
  2018, Proceedings, pp. 57--67 (11 2018)

\bibitem{rip_methods_Duarte_2023}
Duarte, J., Dias, P., Moreira, J.: Approximating the evolution of rotating
  moving regions using bezier curves. International Journal of Geographical
  Information Science  \textbf{37}(4),  839--863 (2023)

\bibitem{Forlizzi_2000}
Forlizzi, L., G\"{u}ting, R.H., Nardelli, E., Schneider, M.: A data model and
  data structures for moving objects databases. In: Proceedings of the 2000 ACM
  SIGMOD International Conference on Management of Data. p. 319–330 (2000)

\bibitem{rip_Heinz_Guting_2020}
Heinz, F., Güting, R.H.: A polyhedra-based model for moving regions in
  databases. International Journal of Geographical Information Science
  \textbf{34}(1),  41--73 (2020)

\bibitem{shape_based_herman}
Herman, G., Zheng, J., Bucholtz, C.: Shape-based interpolation. IEEE Computer
  Graphics and Applications  \textbf{12}(3),  69--79 (1992)

\bibitem{AE_Hinton}
Hinton, G.E., Zemel, R.S.: Autoencoders, minimum description length and
  helmholtz free energy. In: Proceedings of the 6th International Conference on
  Neural Information Processing Systems. p. 3–10. NIPS'93 (1993)

\bibitem{hausdorff_distance}
Huttenlocher, D., Klanderman, G., Rucklidge, W.: Comparing images using the
  hausdorff distance. IEEE Transactions on Pattern Analysis and Machine
  Intelligence  \textbf{15}(9),  850--863 (1993)

\bibitem{VAE_Kingma2013}
Kingma, D.P., Welling, M.: Auto-encoding variational bayes  (12 2013)

\bibitem{Lee_2002}
Lee, T.Y., Lin, C.H.: Feature-guided shape-based image interpolation. IEEE
  Transactions on Medical Imaging  \textbf{21}(12),  1479--1489 (2002)

\bibitem{Mckenney_mov_regions}
Mckenney, M., Frye, R.: Generating moving regions from snapshots of complex
  regions. ACM Trans. Spatial Algorithms Syst.  \textbf{1}(1) (jul 2015)

\bibitem{Pyspatiotemporalgeom}
McKenney, M., Nyalakonda, N., McEvers, J., Shipton, M.: Pyspatiotemporalgeom: A
  python library for spatiotemporal types and operations. In: Proceedings of
  the 24th ACM SIGSPATIAL International Conference on Advances in Geographic
  Information Systems. SIGSPACIAL '16 (2016)

\bibitem{rip_methods_Mckennney_2015}
Mckennney, M., Frye, R.: Generating moving regions from snapshots of complex
  regions  \textbf{1}(1) (2015)

\bibitem{latent_space_interpol_LU}
Mi, L., He, T., Park, C.F., Wang, H., Wang, Y., Shavit, N.: Revisiting
  latent-space interpolation via a quantitative evaluation framework. ArXiv
  \textbf{abs/2110.06421} (2021)

\bibitem{moreira_2016}
Moreira, J., Dias, P., Amaral, P.: Representation of continuously changing data
  over time and space: Modeling the shape of spatiotemporal phenomena. In: IEEE
  12th International Conference on e-Science (e-Science). pp. 111--119 (2016)

\bibitem{AE_interpol}
Oring, A., Yakhini, Z., Hel-Or, Y.: Autoencoder image interpolation by shaping
  the latent space. In: Proceedings of the 38th International Conference on
  Machine Learning. vol.~139, pp. 8281--8290 (2021)

\bibitem{VAE_Rezende2014}
Rezende, D.J., Mohamed, S., Wierstra, D.: Stochastic backpropagation and
  approximate inference in deep generative models. p. II–1278–II–1286
  (2014)

\bibitem{artigo_seg_burned}
Ribeiro, T.F.R., Silva, F., Costa, R.L.C.: Burned area semantic segmentation: a
  novel dataset and evaluation using convolutional networks [manuscript
  submitted for publication] (2023)

\bibitem{Ronneberger2015}
Ronneberger, O., Fischer, P., Brox, T.: U-net: Convolutional networks for
  biomedical image segmentation. In: Medical Image Computing and
  Computer-Assisted Intervention--MICCAI 2015: 18th International Conference,
  Munich, Germany, October 5-9, 2015, Proceedings, Part III 18. pp. 234--241.
  Springer (2015)

\bibitem{shape_based_schenk}
Schenk, A., Prause, G., Peitgen, H.O.: Efficient semiautomatic segmentation of
  3d objects in medical images. In: Medical Image Computing and
  Computer-Assisted Intervention -- MICCAI 2000. pp. 186--195 (2000)

\bibitem{cVAE_2015}
Sohn, K., Yan, X., Lee, H.: Learning structured output representation using
  deep conditional generative models. In: Proceedings of the 28th International
  Conference on Neural Information Processing Systems - Volume 2. p.
  3483–3491. NIPS'15, MIT Press (2015)

\bibitem{Tossebro_Guting_2001}
T{\o}ssebro, E., G{\"u}ting, R.H.: Creating representations for continuously
  moving regions from observations. In: Advances in Spatial and Temporal
  Databases. pp. 321--344 (2001)

\end{thebibliography}

\end{document}